\documentclass[preprint,prl]{revtex4-1}

\usepackage{graphicx}
\usepackage{color}
\usepackage{epstopdf}
\usepackage{amsmath,amssymb,bm,mathrsfs}

\newcommand{\Brho}{\boldsymbol\rho}

\newcommand{\lap}{\nabla^{2}}

\newcommand{\grad}{\nabla}

\newcommand{\Br}{{\bf r}}

\newcommand{\Bp}{{\bf p}}

\newcommand{\Bk}{{\bf k}}

\renewcommand{\k}{{\bf \hat k}}
\newcommand{\z}{{\bf \hat z}}

\newcommand{\BR}{{\bf R}}

\renewcommand{\BR}{{\bf R}}

\newcommand{\ket}[1]{\left| #1 \right\rangle}
\newcommand{\bra}[1]{\left\langle #1 \right|}

\newcommand{\Epos}{\widehat E^{+}}
\newcommand{\Eneg}{\widehat E^{-}}

\begin{document}

\title{Transport of Entanglement}
\author{Manabu Machida}
\affiliation{Department of Mathematics, University of Michigan, Ann Arbor, MI 48109}
\email{mmachida@umich.edu}
\author{Vadim A. Markel}
\affiliation{Department of Radiology, University of Pennsylvania, Philadelphia, PA 19104}
\email{vmarkel@mail.med.upenn.edu}
\author{John C. Schotland}
\affiliation{{}Department of Mathematics, University of Michigan, Ann Arbor, MI 48109}
\email{schotland@umich.edu}
\author{\emph{Dedicated to Emil Wolf on the occasion of his 90th birthday}}

\date{\today}

\begin{abstract}
We consider the propagation of two-photon light in a random medium. We show that the Wigner distribution of the two-photon wave function obeys an equation that is analogous to the radiative transport equation for classical light. Using this result, we predict that the entanglement of a photon pair is destroyed with propagation. 
\end{abstract}
 
\maketitle
 
The propagation of light in disordered media, including clouds, colloidal suspensions and biological tissues, is generally considered within the framework of classical optics~\cite{vanRossum_1999}. However, recent experiments have demonstrated the existence of novel effects in multiple light scattering, in which the quantized nature of the electromagnetic field is evident. These include (i) the transport of quantum noise through random media~\cite{Lodahl_2005_1} (ii) the observation of spatial correlations in multiply-scattered squeezed light~\cite{Smolka_2009, Smolka_2012} (iii) the measurement of two-photon speckle patterns and the observation of non-exponential statistics for two-photon correlations~\cite{Peeters_2010,Pires_2012} and (iv) the finding that interference survives averaging over disorder and is manifested as  photon correlations, exhibiting both antibunching and anyonic symmetry~\cite{Smolka_2011,vanExter_2012}. Thus, there is an interplay between quantum interference and interference due to multiple scattering that is of fundamental interest~\cite{Lodahl_2005_2,Lodahl_2006_1,Lodahl_2006_2,Patra_1999,Beenakker_2009,Tworzydlo_2002,Beenakker_1998} and considerable applied importance.  Indeed, applications to spectroscopy~\cite{ Skipetrov_2007}, two-photon imaging~\cite{Klyshko_1988,Strekalov_1995,Abouraddy_2001,Abouraddy_2004,Gatti_2004,Scarcelli_2004, Scarcelli_2006,Erkmen_2008,DAngelo_2005,Schotland_2010} and quantum communication~\cite{Moustakas_2000,Skipetrov_2008,Shapiro_2009} have been reported.

In the multiple-scattering regime, the radiative transport equation (RTE) governs the propagation of light in random media~\cite{vanRossum_1999}. The RTE is a conservation law that accounts for gains and losses of electromagnetic energy due to scattering and absorption. The physical quantity of interest is the specific intensity $I(\Br,\k)$, defined as the intensity at the position $\Br$ in the direction $\k$. The specific intensity obeys the RTE
\begin{equation}
\label{RTE}
\k \cdot \grad_\Br I + \mu_a I =\mu_s \int d^2 k'\left[p(\k',\k)I(\Br,\k')-p(\k,\k')I(\Br,\k)\right] \  ,
\end{equation}
which we have written in its stationary form. Here $\mu_a$ and $\mu_s$ are the absorption and scattering coefficients of the medium and $p$ is the phase function. We note that although the RTE is often viewed as phenomenological, it is derivable from the scattering theory of electromagnetic waves in a random medium~\cite{vanRossum_1999,Ryzhik_1996,Wolf_1976}.

The propagation of two-photon light is generally considered either in free space or, in some cases, with account of diffraction~\cite{Saleh_2000,Abouraddy_2002}. However, understanding the interaction of light with matter is central to applications in both imaging and quantum information. In this Letter, we consider the propagation of two-photon light in a random medium. We show that the averaged Wigner distribution of the two-photon wave function obeys an equation that is analogous to the RTE. Using this result, we characterize the loss of entanglement of a photon pair upon propagation in a random medium. In this sense, our work builds on the well-known duality between partially coherent and partially entangled light, in which loss of entanglement is dual to the gain in coherence with propagation~\cite{Saleh_2000,Abouraddy_2002}.

We begin by recalling some important facts about two-photon light. We consider the two-photon state $\ket{\psi}$ and define the  second-order coherence function as the normally ordered expectation of field operators:
\begin{eqnarray}
\Gamma^{(2)}(\Br_1,t_1;\Br_2,t_2) = \bra{\psi}\Eneg(\Br_1,t_2)\Eneg(\Br_2,t_2)\Epos(\Br_2,t_2)\Epos(\Br_1,t_1)\ket{\psi}
\  , 
\end{eqnarray}
where $\Eneg$ and $\Epos$ are the negative- and positive-frequency components of the electric-field operator with $\Eneg = [\Epos]^{\dag}$.
In a material medium with dielectric permittivity $\varepsilon$, the field operator $\Epos$ obeys the wave equation~\cite{Glauber_1991,Scheel_1999}
\begin{equation}
\label{wave_eqn}
\lap \Epos - \frac{\varepsilon(\Br)}{c^2} \frac{\partial^2 \Epos}{\partial t^2} = 0 \ .
\end{equation}
Here the medium is taken to be nonabsorbing, so that $\varepsilon$ is purely real. 

The quantity $\Gamma^{(2)}$ is proportional to the probability of detecting one photon at $\Br_1$ and a second photon at $\Br_2$ and can be measured in a Hanbury-Brown--Twiss interferometer~\cite{Mandel_Wolf}. For the two-photon state $\ket{\psi}$, it can be seen that $\Gamma^{(2)}$ factorizes~\cite{Rubin_1996,Saleh_2000} as follows:
\begin{widetext}
\begin{eqnarray}
\Gamma^{(2)}(\Br_1,t_1;\Br_2,t_2) &=& \sum_n \bra{\psi}\Eneg(\Br_1,t_1)\Eneg(\Br_2,t_2)\ket{n}\bra{n}\Epos(\Br_2,t_2)\Epos(\Br_1,t_1)\ket{\psi} \\
&=& \bra{\psi}\Eneg(\Br_1,t_1)\Eneg(\Br_2,t_2)\ket{0}\bra{0}\Epos(\Br_2,t_2)\Epos(\Br_1,t_1)\ket{\psi} \\ 
&=&|\Phi(\Br_1,t_1;\Br_2,t_2)|^2 \ .
\end{eqnarray}
\end{widetext}
Here $\{\ket{n}\}$ denotes a complete set of states and the two-photon
probability amplitude $\Phi$ is defined by
\begin{equation}
\Phi(\Br_1,t_1;\Br_2,t_2) = \bra{0}\Epos(\Br_1,t_1)\Epos(\Br_2,t_2)\ket{\psi} \ .
\end{equation}
Evidently, $\Phi$ satisfies the pair of wave equations
\begin{eqnarray}
\label{phi_1}
\lap_{\Br_j} \Phi - \frac{\varepsilon(\Br_j)}{c^2}\frac{\partial^2 \Phi}{\partial t_j^2} = 0 \ , \quad j=1,2 \ .
\end{eqnarray}
which  follow from the fact that $\Epos$ obeys the wave equation (\ref{wave_eqn}). We note that (\ref{phi_1}) is the analog the Wolf equations for two-photon light~\cite{Saleh_2005}. We will find it convenient to introduce the Fourier transform of the probability amplitude $\Phi$, which is given by
\begin{equation}
\widetilde\Phi(\Br_1,\omega_1;\Br_2,\omega_2)= \int dt_1 dt_2 e^{i(\omega_1 t_1 + \omega_2 t_2)} \Phi(\Br_1,t_1;\Br_2,t_2) \ .
\end{equation}
Eq.~(\ref{phi_1}) then becomes
\begin{eqnarray}
\label{phi_1_helmholtz}
\lap_{\Br_j} \widetilde\Phi + k_j^2\varepsilon(\Br_j) \widetilde\Phi = 0 \ , \quad j=1,2 \ .
\end{eqnarray}
where $k_j=\omega_j/c$. 
It is important to note that if $\widetilde\Phi$ factorizes into a product of two functions which depend upon $\Br_1$ and $\Br_2$ separately, then the two-photon state $\ket{\psi}$ is not entangled. In contrast, a \emph{fully entangled state} is not separable and corresponds to
$\widetilde\Phi(\Br_1,\Br_2) \propto \delta(\Br_1-\Br_2)$.

We now consider the Wigner distribution of $\widetilde\Phi$ which is defined by
\begin{equation}
\label{wigner}
W(\Br,\Bk) = \int d^3r' e^{i\Bk\cdot\Br'}
\widetilde\Phi(\Br-\Br'/2,\omega_1;\Br+\Br'/2,\omega_2) \ .
\end{equation}
Upon subtracting (\ref{phi_1_helmholtz}) for $j=1,2$ and changing variables according to
$\Br_1 = \Br - \Br'/2,  \ \Br_2 = \Br + \Br'/2$,
we see that $W$ obeys the equation
\begin{widetext}
\begin{equation}
\label{liouville_eq}
\Bk \cdot \grad_\Br W + \frac{i}{2} \int \frac{d^3p}{(2\pi)^3} 
e^{-i\Bp\cdot\Br} \widetilde\varepsilon(\Bp)\left[k_1^2 W(\Br,\Bk+ \Bp/2)-k_2^2 W(\Br,\Bk - \Bp/2)\right] = 0 \ .
\end{equation}
\end{widetext}
We note that~(\ref{liouville_eq}) is an \emph{exact} result which describes the propagation of the Wigner distribution for two-photon light in a material medium.

We now proceed to derive the RTE for two-photon light.
To this end, we consider a statistically homogeneous random medium and assume that the susceptibility $\eta$ is a Gaussian random field with correlations
$\langle \eta(\Br) \rangle = 0$,
$\langle \eta(\Br)\eta(\Br')\rangle = C(|\Br-\Br'|)$. Here $\eta$ is related to the dielectric permittivity by 
$\varepsilon = 1+4\pi \eta$, $C$ is the two-point correlation function and $\langle \cdots \rangle$ denotes statistical averaging. Let $L$ denote the propagation distance of the field and $\xi$ the correlation length over which $C$ decays at large distances. We introduce a small parameter $\epsilon = 1/(k_0L) \ll 1$ and suppose that the fluctuations in $\eta$ are sufficiently weak that $C$ is of the order $O(\epsilon)$ and $\xi/L=O(\epsilon)$. We then rescale the spatial variables according to $\Br_1\to \Br_1/\epsilon$, $\Br_2\to \Br_2/\epsilon$ and define the scaled  two-photon probability amplitude $\Phi_{\epsilon}(\Br_1,\omega_1;\Br_2,\omega_2)=\widetilde\Phi(\Br_1/\epsilon,\omega_1;\Br_2/\epsilon,\omega_2)$, so that (\ref{phi_1_helmholtz}) becomes
\begin{eqnarray}
\label{scaled}
\epsilon^2\lap_{\Br_j} \Phi_{\epsilon} +  k_j^2\Phi_{\epsilon}= -4\pi k_j^2 \sqrt{\epsilon} \eta\left(\Br_j/\epsilon\right) \Phi_{\epsilon}\ , \quad j=1,2 \ .
\end{eqnarray}
where we have introduced a rescaling of $\eta$ to be consistent with the assumption that the fluctuations are of size $O(\epsilon)$. If we denote by $W_\epsilon$ the Wigner distribution of $\Phi_\epsilon$, defined according to (\ref{wigner}), then 
(\ref{liouville_eq}) becomes
\begin{equation}
\label{liouville_eq_eps}
\Bk \cdot \grad_\Br W_\epsilon + \frac{i}{2\epsilon}\left(k_1^2 - k_2^2\right) W_\epsilon + \frac{1}{\sqrt{\epsilon}} \mathscr{L} W_\epsilon = 0\ ,
\end{equation}
where
\begin{eqnarray}
\nonumber
\mathscr{L} W_\epsilon &=&
2\pi i \int \frac{d^3p}{(2\pi)^3} 
e^{-i\Bp\cdot\Br/\epsilon} \widetilde\eta(\Bp)\Big[k_1^2 W_\epsilon(\Br,\Bk+ \Bp/2) \\
&& \quad\quad\quad -k_2^2 W_\epsilon(\Br,\Bk - \Bp/2)\Big]  \ .
\end{eqnarray}
We now consider the asymptotics of the Wigner distribution in the homogenization limit $\epsilon \to 0$. This corresponds to the regime of  weak fluctuations. Following standard procedures~\cite{Ryzhik_1996}, we introduce a two-scale expansion for $W_\epsilon$ of the form
\begin{equation}
W_\epsilon(\Br,\BR,\Bk) = W_0(\Br,\BR,\Bk) + \sqrt{\epsilon} W_1(\Br,\BR,\Bk) + \epsilon W_2(\Br,\BR,\Bk) + \cdots \ ,
\end{equation}
where $\BR=\Br/\epsilon$ is a fast variable. Next, we suppose that $\mu =(k_1^2 - k_2^2)/(2k\epsilon) = O(1)$, which corresponds to working in the high-frequency regime. By averaging over the fluctuations on the fast scale,
it can be seen that $\langle W_0 \rangle$, which we denote by $\mathcal{I}$, obeys the equation
\begin{equation}
\label{2photonRTE}
\k \cdot \grad_\Br \mathcal{I}(\Br,\k) + (\mu_a + \mu_s) I(\Br,\k) =  \mu_s \int d^2k' f(\k,\k')\mathcal{I}(\Br,\k') \  .
\end{equation}
Here the absorption coefficient $\mu_a$, scattering coefficient $\mu_s$ and scattering kernel $f$ are defined by
\begin{eqnarray}
\mu_a &=& (k_1^4-k_2^4) \int  \widetilde C(k(\k-\k'))d^2k' + i \mu \ ,\\
\mu_s &=& k_2^2(k_1^2+k_2^2)\int  \widetilde C(k(\k-\k'))d^2k'   \ ,
\end{eqnarray}
where $f(\k,\k') =\widetilde C(k(\k-\k'))$ is normalized so that $\int f(\k,\k')d^2k' =1$ for all $\k$. We note that this normalization is consistent with the statistical homogeneity of the random medium since $\widetilde C(k(\k-\k'))$ depends only upon the quantity $\k\cdot\k'$. We will refer to~(\ref{2photonRTE}) as the two-photon RTE. Evidently, (\ref{2photonRTE}) is the analog of the classical RTE. However, the physical interpretation of (\ref{2photonRTE}) requires some care. In contrast to the specific intensity, the quantity $\mathcal{I}$ is not real-valued and is not directly measurable. Nevertheless, by inversion of the Fourier transform (\ref{wigner}), we find that $\mathcal{I}$ is related to the average two-photon probability amplitude by means of the formula
\begin{equation}
\langle\widetilde\Phi(\Br_1,\Br_2)\rangle =
\int \frac{d^3 k}{(2\pi)^3} e^{i\Bk\cdot (\Br_1-\Br_2)} \mathcal{I}\left(\frac{\Br_1+\Br_2}{2},\k\right) \ , 
\label{inverse}
\end{equation}
where the dependence on the frequencies $\omega_1$ and $\omega_2$ has not been indicated.

We now explore some physical consequences of the two-photon RTE. In particular, we examine the propagation of entanglement. We begin with the case of a \emph{deterministic} medium in which the permittivity $\varepsilon$ is constant.
We consider the half-space $z\ge 0$ and assume that the two-photon Wigner distribution $\mathcal I_0$ is specified on the disk of radius $a$ in the plane $z=0$ in the ingoing direction. That is,
\begin{equation}
{\mathcal I_0}(\Br,\k) = 
\begin{cases}
A\delta(k-k_0) \quad {\rm if}\quad  \k\cdot\z >0 \ {\rm and} \ |\Brho|\le a  \ \\
0 \quad {\rm otherwise} \ ,
\end{cases}
\end{equation}
where $A$ is constant and $\Brho$ is the transverse coordinate in the $z=0$ plane.
Making use of~(\ref{inverse}), it is readily seen that 
\begin{equation}
\widetilde\Phi(\Brho_1,0;\Brho_2,0)=
\begin{cases}
 2\pi k_0^2 A \frac{\sin\left(k_0|\Brho_1 - \Brho_2|\right)}{(k_0|\Brho_1 - \Brho_2|}
 \quad {\rm if} \quad  |\Brho_{1,2} |\le a  \\
0 \quad {\rm otherwise} \  ,
\end{cases}
\end{equation}
which corresponds to a transversely entangled two-photon state. To propagate $\mathcal I$ into the $z>0$ half-space, we make use of the formula
\begin{equation}
\label{int_rep}
{\mathcal I}(\Br,\k) = \int d^2k' \int_{z'=0} d^2r' \; \z\cdot \k'
G(\Br,\k;\Br',\k') {\mathcal I}_0(\Br',\k')  \  .
\end{equation}
Here $G$ is the Green's function for the two-photon RTE (\ref{liouville_eq}), which is given by
\begin{equation}
G(\Br,\k,\Br',\k') = \frac{1}{|\Br-\Br'|^2} \delta(\k-\k')\delta\left(\k-\frac{\Br-\Br'}{|\Br-\Br'|}\right) \ .
\end{equation}
We can now compute the two-photon probability amplitude $\widetilde\Phi$. For simplicity, we assume that $k_1=k_2=k_0$ and that the points of observation $\Br_1$ and $\Br_2$ are on-axis, with $\Br_1=\Br_2=(0,z)$. Carrying out the integrations in (\ref{int_rep}) and making use of (\ref{inverse}) we find that 
\begin{equation}
\widetilde\Phi(0,z;0,z) = A\left(\frac{k_0}{2\pi}\right)^2 \tan^{-1}\left(\frac{a}{2z} \right) \ .
\end{equation}
It can be seen that the entanglement of the photon pair is destroyed with propagation. We note that the diagonal part of the coherence function $\Gamma^{(2)}(\Br,\Br)= |\widetilde\Phi(\Br,\Br)|^2$ is proportional to the probability of two-photon absorption at the point $\Br$. Next, we consider the case of a \emph{random} medium. 
At first, we will make use of the diffusion approximation (DA) to the RTE, which is widely used in applications. The DA neglects the angular dependence of the Green's function. It holds in the limit of strong scattering and at large distances from the source. Within the accuracy of the DA, the Green's function
for the RTE is given by $G(\Br,\Br')  = \exp(-\kappa|\Br-\Br'|)/(4\pi D |\Br-\Br'|)$.
Here $\kappa=\sqrt{\mu_a/D}$, $D=1/[3(\mu_a + (1-g)\mu_s)]$ and  $g=\int \k\cdot\k' f(\k,\k')d^2k'$.
Carrying out the integrations in (\ref{inverse}) and ({\ref{int_rep}), we find tha
the average two-photon probability amplitude is given by
\begin{eqnarray}
\label{entanglement_DA}
&&
\langle\widetilde\Phi(\Br_1,\Br_2)\rangle = \frac{a A k_0}{2D(2\pi)^2} \frac{\sin{(k_0|\Br_1-\Br_2|)}}{|\Br_1-\Br_2|} \\ 
&&\times\int_0^\infty \frac{dq}{\sqrt{q^2 + \kappa^2}}  J_1(q a)J_0(q|\Brho_1+\Brho_2|/2) e^{-\sqrt{q^2+\kappa^2} (z_1+z_2)/2} \ ,
\nonumber
\end{eqnarray}
where $\Br=(\Brho,z)$.  In the on-axis configuration,  we find that
\begin{equation}
\label{entanglement_DA_onaxis}
\langle\widetilde\Phi(0,z;0,z)\rangle = \frac{ A k_0}{2D(2\pi)^2} \left[\sqrt{z^2+ a^2}-z \right] \ .
\end{equation}
As above, the entanglement of the photon pair is destroyed with propagation.
\begin{figure}[t]     
\centering
\includegraphics[width=4.5in]{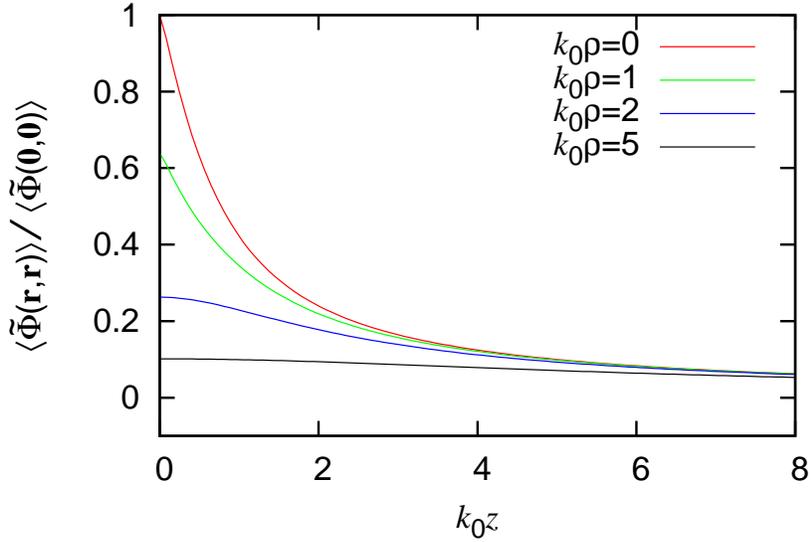} 
\caption{(Color Online) Dependence of $\langle\widetilde\Phi\rangle$ on the distance of propagation $z$ for different off-axis distances $\rho$. The scattering kernel $f$ was taken to be of the Henyey-Greenstein form with $g=0.9$, 
$\mu_a=0$ and $\mu_s = 100k_0$.}
\label{fig1}
\end{figure}
Finally, we consider the propagation of entanglement in the transport regime. The Green's function for the two-photon RTE can be obtained using the method of rotated reference frames~\cite{Markel_2004,Panasyuk_2006,Machida_2010}. The calculation of the average two-photon probability amplitude is presented in the supplementary material. In Fig.~1 we plot the $z$-dependence of $\langle\widetilde\Phi(\Br,\Br)\rangle$ for various values of the off-axis distance $\rho$ with $k_1=k_2=k_0$. Once again, the entanglement of the photon pair is lost with propagation. 

We close with a few remarks. (i) It is possible to derive the analog of the RTE for single photons. Not surprisingly, this equation has the form of the classical RTE (\ref{RTE}). We plan to present 
(ii) Although in our model the electromagnetic field is quantized, the interaction of the field with the scattering medium is treated classically. It would be of interest to extend our results to the case in which the medium consists of a collection of two- or three-level atoms. In this manner, it should (in principle) be possible to understand the transfer of entanglement from the field to the medium~\cite{Berman_2007}. Evidently, the calculations that we have presented do not account for this effect, since we have taken a macroscopic approach to the quantization of the field~\cite{Glauber_1991,Scheel_1999}.
(iii) Finally, applications to imaging and communication theory may be envisioned. In the former case, there has been extensive use of the classical RTE for imaging in random media. It may be anticipated that experiments with two-photon light may enjoy some advantages, as has been suggested for the case of quantum optical coherence tomography~\cite{Nasr_2003,Teich_2012}. In the latter case, there has been considerable interest in the use of quantum states of light for communication~\cite{Moustakas_2000,Shapiro_2009,Yuan_2010}. It would be of interest to understand the effect of a random medium, such as the atmosphere, on the capacity of quantum information systems~\cite{Skipetrov_2008}.

We are grateful to Paul Berman, Scott Carney, Roberto Merlin and Ted Norris for valuable discussions. This work was supported in part by the NSF grants DMR--1120923, DMS--1115574 and DMS--1108969.

\end{document}